\documentclass [11pt]{article}
\title{Can One-Way Light Speed be Measured?\\Comment on Greaves et al, Am. J. Phys. 77(10), 894-896 (2009)}
\author{Robert D. Klauber\\1100 University Manor Dr., 38B \\Fairfield, Iowa 52556\\rklauber(AT)netscape.net}

\date{January 22, 2009}

\usepackage{graphicx}
\usepackage{longtable}

\begin{document}
\maketitle

Greaves et al\cite{Greaves:2009} claim they devised an experiment to 
measure the one-way speed of light, something the vast majority of 
specialists consider wholly dependent on the particular synchronization 
convention chosen (with an infinitude of possibilities), and therefore not 
measurable in any absolute sense.\cite{There:1998} I argue here on the side 
of the majority.

\section{THE AMBIGUITY IN ONE-WAY LIGHT SPEED}
\label{sec:mylabel1}
One-way light speed shares an interdependence with synchronization 
convention, i.e., the arbitrary setting of separate clocks at the photon 
emission and reception locations. The time difference between the first 
clock reading at the emission event and the second at the reception event is 
used with the known distance between clocks to calculate the speed. 
Different clock settings mean different one-way speeds. Round trip speed, on 
the other hand, whose emission and reception times are found using the same 
clock, does not require synchronization of distant clocks, is not subject to 
convention, and always (in vacuum) equals the universal constant $c$. If, for a 
given synchronization convention, the outgoing one-way speed is less than 
$c$, then the return one-way speed is greater than $c$, in such a way as to make 
the average round trip speed $c$. This conclusion is valid for any type of 
round trip path (back and forth in a straight line, loops, etc.)

One can either set distant clocks arbitrarily and use them to determine 
one-way light speed, or conversely, assume a particular one-way light speed 
and use it to set distant clocks.\cite{Start:1} In Einstein 
synchronization, the most common type, one assumes a one-way speed of $c$ and 
uses that to set distant clocks. In other synchronizations, one assumes a 
different one-way speed.\cite{The:1}

Greaves et al's purpose was to determine one-way light speed using the 
single clock at the detector, by varying the distance of one-way light 
travel and measuring the time variation on that single clock. Since only a 
single clock is used, the one-way speed would presumably be convention 
independent. Though this may seem reasonable at first, unfortunately it does 
not work, as explained below.

\section{WHAT THE EXPERIMENT MEASURED}
Fig. 1 is equivalent to, though slightly embellished from, the first figure 
in Greaves et al. The path from B to D is effectively horizontal, though, to 
make the figure legible, it is drawn as slightly inclined. We assume that 
the one-way speed of light from right to left in the horizontal direction 
equals $c^{- }<c$. This assumption means that if there were clocks at the 
sensor locations $L_{3}$ and $L_{2}$, they would be set relative to the 
detector clock at point $C$ such that they yield a one-way speed $c^{-}$.

The length of the cable of Fig. 1 is 23.73 m, and the time delay cited for 
an electromagnetic signal passing from one end of the cable to the other is 
79 ns. In other communication, the lead author implied this was derived by 
taking half the time for a back and forth measurement, with reflection off 
the far end of the cable, using a single clock, though it may have simply 
been assumed. This yields a round trip signal speed in the cable of $c$ 
$\approx $ 3.00 $\times $ 10$^{8}$ m/s, essentially the same as that in 
air.\cite{This:1} It follows that the one-way speed of light in the cable 
would essentially equal the one-way speed in air.

If $c^{- }<c$, then as the cable from the sensor to the detector is looped 
to bring the sensor that was at \textbf{\textit{L}}$_{3}$ to 
\textbf{\textit{L}}$_{2}$, the time for the signal to pass the entire length 
of the cable does not, as claimed, stay the same. This is because, in the 
loop(s), the signal is traveling a round trip path, with average speed $c$, 
whereas elsewhere in the cable, it is traveling a one-way path with speed 
$c^{-}$.\cite{Any:1} For a straight cable aligned horizontally in Fig. 1 
(lower configuration), the speed along the entire length would be $c^{-}$.

The contribution from the $L_{3}$ to $L_{2}$ section of the path to the 
measured time on the clock at detector C is the same, whether it is the 
visible light signal in air or the electromagnetic signal in the cable that 
is passing between $L_{3}$ and $L_{2}$. The closed loop length $L_{L}$ equals 
the distance between locations $L_{3}$ and $L_{2}$, which is the additional 
distance in the total path from the case with the sensor at 
\textbf{\textit{L}}$_{3}$ (lower configuration) to that with the sensor at 
\textbf{\textit{L}}$_{2}$ (upper configuration). Hence, the only measured 
time difference between the two sensor location cases is from the closed 
loop part of the cable.

\begin{figure}[htbp]
\centerline{\includegraphics[width=4.36in,height=2.29in]{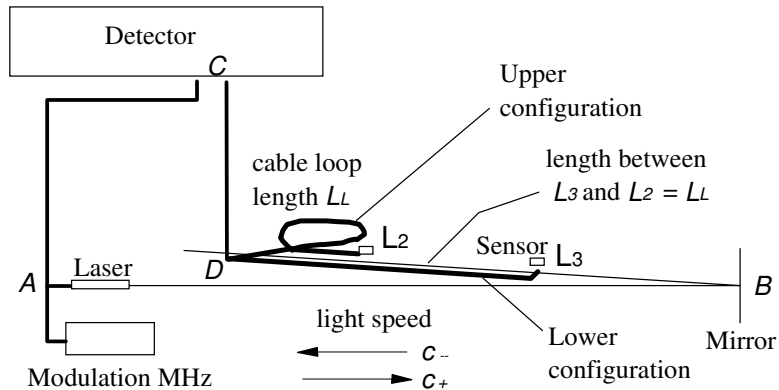}}
\caption{Sensor Location First at {\textit{L}}$_{3}$, then at {\textit{L}}$_{2}$.}
\label{fig1}
\end{figure}

And there, the electromagnetic signal speed equals $c$, meaning the experiment 
actually measures round trip speed in the loop, not one-way speed between 
$L_{3}$ and $L_{2}$.

\section*{Appendix I. Cable Signal Speed of 2/3 c}
For a cable round trip signal speed of 2/3 c, start by taking $\Delta 
T_{base}$ as the time for the signal to pass along the cable from $L_{2}$ to 
point $D$, where the cable turns vertical in Fig. 1. The time for the signal to 
pass from $L_{3}$ to $D$ in the lower configuration of Fig. 1, where $c_{coax}^- 
$ is one way speed from right to left in the cable, is
\begin{equation}
\label{eq1}
T_{lower} =\Delta T_{base} +\frac{L_L }{c_{coax}^- }.
\end{equation}
The time for the signal to pass from $L_{3}$ to $D$ in the upper configuration 
is
\begin{equation}
\label{eq2}
T_{upper} =\Delta T_{base} +\frac{L_L }{\textstyle{2 \over 3}c}+\frac{L_L 
}{c^-}.
\end{equation}
The difference is
\begin{equation}
\label{eq3}
T_{upper} -T_{lower} =\frac{L_L }{\textstyle{2 \over 3}c}+\frac{L_L 
}{c^-}-\frac{L_L }{c_{coax}^- }.
\end{equation}
If one uses Einstein synchronization, where $c^{-}=c$, and $c_{coax}^- 
=\textstyle{2 \over 3}c$, then one has
\begin{equation}
\label{eq4}
T_{upper} -T_{lower} =\frac{L_L }{c},
\end{equation}
where the time difference in (\ref{eq4}) and $L_{L}$ were measured in the experiment.

Synchronizations other than Einstein's change nothing that would be measured 
in (\ref{eq3}), since the time difference in (\ref{eq3}) is on a single clock, and this is 
unaffected by the synchronization choice for distant clocks. For any such 
other synchronization, $c^{-}$ would be known, and the constitutive value 
$c_{coax}^- $ would be determined by experiment. That experiment could be 
this one. Simply use the same experimental values and solve (\ref{eq3}) for 
$c_{coax}^- $. In any synchronization choice, the time difference in (\ref{eq3}) and 
(\ref{eq4}), as well as $L_{L}$, are the same, so (\ref{eq4}) holds for all synchronizations.

\section*{Appendix II. R\"{o}mer Experiment}
Some may consider the famous R\"{o}mer experiment, which measured the speed 
of light via changes in light transmission time from Jupiter and its moons 
to be a one-way light speed determination. This effectively entails a time 
difference measured via a single clock on the Earth as this clock moves to 
different positions in the Earth's orbit.

However, as shown in Ref. \cite{There:1998}, the setting on a clock 
that is slowly transported relative to the speed of light (as the Earth 
clock is), will, upon arriving at a distant location, read the same time as 
a clock at that location that had been synchronized via the Einstein 
convention. That is, regardless of the synchronization convention used (and 
thus regardless of the one-way speed of light), the slowly transported clock 
will be set as though it had been Einstein synchronized.

Thus, any measurement with such a slowly transported clock will show $c$ for 
the one-way speed of light, as it is for Einstein synchronization. Hence, 
the R\"{o}mer measurement does not measure one-way light speed in the usual 
sense.

There has, however, been more than a modicum of philosophical debate over 
whether slowly transported clocks should be taken as nature's edict that an 
absolute synchronization (Einstein's) exists. The present consensus is that 
it should not.


\begin{thebibliography}{3}
\bibitem{This:1} This result seems strange, as signal speed in a coaxial cable is typically 2/3 that in air. See Appendix I for an analysis of that more realistic case.
\bibitem{Any:1} Any folding back of the cable not in a loop would have the same net result. For example, an S curve would, from right-to-left, be a twice back and forth passage of the signal.
\end{thebibliography}
\end{document}